\documentclass[10pt,conference]{IEEEtran}

\usepackage{filecontents,lipsum}  % http://ctan.org/pkg/lipsum
\usepackage{longtable}
\usepackage{balance}
\usepackage{boxedminipage}
\usepackage[linesnumbered,ruled,vlined]{algorithm2e}
\usepackage{newtxmath}
\usepackage{url}

\usepackage{wrapfig}
\usepackage{rotating}
\usepackage{multicol, multirow, makecell}% http://ctan.org/pkg/multirow
\usepackage{tabularx, tabulary}
\usepackage{hhline}     % http://ctan.org/pkg/hhline
\usepackage{xspace}
\usepackage{xcolor}
\usepackage{array}
\usepackage{zi4}
\usepackage{ragged2e}   % for '\RaggedRight' macro (allows hyphenation)
\usepackage{newfloat}
\DeclareFloatingEnvironment[name={Algorithm}]{algfigure}

\usepackage{graphicx}  % package to use graphics
\usepackage[]{subfig}
\usepackage[utf8]{inputenc}
\usepackage[T1]{fontenc}

\usepackage{booktabs}
\usepackage{graphicx}
\usepackage{color,xcolor}
\let\subparagraph\relax

\usepackage{etoolbox}
\usepackage{stfloats}
\usepackage{afterpage}

\usepackage{adjustbox}

\definecolor{dkgreen}{rgb}{0,0.6,0}
\definecolor{mauve}{rgb}{0.58,0,0.82}
\definecolor{jcolor}{rgb}{0.5,0,0.5}

\definecolor{brickred}{RGB}{191,85,85}
\definecolor{ltblue}{RGB}{40,139,228}
\definecolor{ycolor}{RGB}{242,139,96}

\definecolor{narratives}{RGB}{191,85,85}
\definecolor{integrated}{RGB}{40,139,228}
\definecolor{original}{RGB}{0,0,0}
\definecolor{darkred}{RGB}{138,0,0}
\definecolor{darkcyan}{RGB}{29,91,95}
\definecolor{mcolour}{RGB}{255,0,0}

\usepackage{xspace}
\newcommand{\data}{\textsc{AndroR2}\xspace}

\newcounter{lesson}[section] % this one for 'lesson learned'
\newcounter{redir}[section] % this one for 'research direction'
\usepackage{fontawesome} % this package allows to use light bulb
\usepackage{ulem}[] % the package for strikethrough
\usepackage{pifont}

\newcommand{\checkmarkwithcross}{\textcolor{red}{\ding{51}}\textsuperscript{\textcolor{red}{\kern-0.5em\tiny\ding{55}}}}

\newcolumntype{P}[1]{>{\raggedright\arraybackslash\hspace{0pt}}p{#1}}
\newcolumntype{C}[1]{>{\centering\arraybackslash\hspace{0pt}}p{#1}}

\usepackage{tikz}
\usetikzlibrary{shapes,arrows}
\usetikzlibrary{backgrounds,positioning}

\DeclareMathAlphabet\mathbfcal{OMS}{cmsy}{b}{n}

\newcolumntype{Y}{>{\RaggedRight\arraybackslash}X}

\usepackage{listings}
\usepackage{color}
\usepackage{soul} % the package for the highlight command
\soulregister\cite7
\soulregister\ref7
\soulregister\emph7
\soulregister\ym7
\soulregister\yw7
\soulregister\ms7
\soulregister\mss7
\soulregister\footnote8
\soulregister\textbf7
\soulregister\textcolor7
\soulregister\guillemotleft0

\definecolor{lstkeywordcolor}{rgb}{0,0,0.5}
\definecolor{lstcommentcolor}{rgb}{0,0.4,0}
\definecolor{lststringcolor}{rgb}{0.5,0,0.5}

\tikzset{
  annotation/.style  = {
    draw,
    rectangle callout,
    callout relative pointer={#1},
    fill=white
    },
}

\usepackage[hidelinks]{hyperref}

\usepackage{textcomp}
\usepackage[T1]{fontenc}

\usepackage[most]{tcolorbox}

\definecolor{barcolor}{RGB}{99,142,198}

\usepackage{collcell}
\usetikzlibrary{calc}
\usetikzlibrary{positioning}
\pgfdeclarelayer{background} % define the background layer
\pgfdeclarelayer{main} % define the main layer, can be omit
\pgfsetlayers{background,main} % define the order of the two layers
\definecolor{barlabelcolor}{RGB}{50,71,150}
\definecolor{scalecolor}{rgb}{1,1,1} % invisible lines (white), used to force scale of figure

\tikzstyle{myBarStyle} = [
    fill=barcolor,
    draw=none
]

\tikzstyle{myDashStyle} = [
    fill=none,
    draw=black,
    dashed,
    thick
]

\tikzstyle{myBarLabelStyle} = [
  font=\LARGE,
  anchor=west, % place text right of anchor point
  minimum height=1em,
  black
]

\tikzstyle{myDashLabelStyle} = [
  myBarLabelStyle,
  black
]

\usepackage{comment}
\usepackage[english]{babel} % For definitions 
\usepackage[htt]{hyphenat}

\usepackage{fixltx2e}

\def\BibTeX{{\rm B\kern-.05em{\sc i\kern-.025em b}\kern-.08em
    T\kern-.1667em\lower.7ex\hbox{E}\kern-.125emX}}

\newcommand{\ie}{i.e.,\xspace}

\newboolean{showcomments}

\setboolean{showcomments}{true}

\ifthenelse{\boolean{showcomments}}
  {\newcommand{\nb}[2]{
    \fbox{\bfseries\sffamily\scriptsize#1}
    {\sf\small$\blacktriangleright$\textit{#2}$\blacktriangleleft$}
   }
   
  }
  {\newcommand{\nb}[2]{}
   
  }

\usepackage{tcolorbox}
\tcbset{%
    bugreportstyle/.style={
        enhanced,
        colback=white,
        colframe=black,
        fonttitle=\itshape,
        colbacktitle=white,
        coltitle=black,
        boxrule=1pt,
        boxsep=0pt,
        left=4pt,
        right=4pt,
        before skip=4pt,
        after skip=0pt,
        separator sign none,
        attach boxed title to top left={%
            yshift*=-\tcboxedtitleheight/2, 
            xshift=5mm},
        boxed title style={colframe=black}
    }
}
\newtcbtheorem{bugreport}{}{bugreportstyle}{number}

\usepackage{fancyvrb}
\usepackage{verbatim}
\usepackage{listings}

\definecolor{delim}{RGB}{20,105,176}
\definecolor{numb}{RGB}{106, 109, 32}
\definecolor{string}{rgb}{0.64,0.08,0.08}

\lstdefinelanguage{json}{
    numbers=none,
    xleftmargin=3pt,
    rulecolor=\color{black},
    showspaces=false,
    showtabs=false,
    breaklines=true,
    postbreak=\raisebox{0ex}[0ex][0ex]{\ensuremath{\color{gray}\hookrightarrow\space}},
    breakatwhitespace=true,
    basicstyle=\ttfamily\small,
    upquote=true,
    morestring=[b]",
    stringstyle=\color{string},
    literate=
     *{0}{{{\color{numb}0}}}{1}
      {1}{{{\color{numb}1}}}{1}
      {2}{{{\color{numb}2}}}{1}
      {3}{{{\color{numb}3}}}{1}
      {4}{{{\color{numb}4}}}{1}
      {5}{{{\color{numb}5}}}{1}
      {6}{{{\color{numb}6}}}{1}
      {7}{{{\color{numb}7}}}{1}
      {8}{{{\color{numb}8}}}{1}
      {9}{{{\color{numb}9}}}{1}
      {\{}{{{\color{delim}{\{}}}}{1}
      {\}}{{{\color{delim}{\}}}}}{1}
      {[}{{{\color{delim}{[}}}}{1}
      {]}{{{\color{delim}{]}}}}{1},
}

\begin{document}

\pagestyle{plain}
\pagenumbering{arabic}

\title{\data: A Dataset of Manually-Reproduced Bug Reports for Android apps}

\author{
\IEEEauthorblockN{Tyler Wendland$^{*}$, Jingyang Sun$^{\dagger}$, Junayed Mahmud,$^{\ddagger}$ S. M. Hasan Mansur$^{\ddagger}$, Steven Huang$^{\dagger}$, \\Kevin Moran$^{\ddagger}$, Julia Rubin$^{\dagger}$, Mattia Fazzini$^{*}$}
\IEEEauthorblockA{
\textit{$^{*}$University of Minnesota, MN, USA}; \href{mailto:}{wendl155@umn.edu}, \href{mailto:}{mfazzini@umn.edu}\\
\textit{$^{\dagger}$University of Bristish Columbia, BC, Canada}; \href{mailto:}{sunjingy@ece.ubc.ca}, \href{mailto:}{huang145@student.ubc.ca}, \href{mailto:}{mjulia@ece.ubc.ca}\\
\textit{$^{\ddagger}$George Mason University, VA, USA}; \href{mailto:}{jmahmud@gmu.edu}, \href{mailto:}{smansur4@gmu.edu}, \href{mailto:}{kpmoran@gmu.edu}
}
}

\maketitle

\begin{abstract}

Software maintenance constitutes a large portion of the software development lifecycle. 
To carry out maintenance tasks, developers often need to understand and reproduce bug reports. 
As such, there has been increasing research activity coalescing around the notion of automating various activities related to bug reporting. 
A sizable portion of this research interest has focused on the domain of mobile apps. 
However, as research around mobile app bug reporting progresses, there is a clear need for a manually vetted and \textit{reproducible} set of real-world bug reports that can serve as a benchmark for future work. 
This paper presents \data: a dataset of 90 manually reproduced bug reports for Android apps listed on Google Play and hosted on GitHub,
systematically collected via an in-depth analysis of 459 reports extracted from the GitHub issue tracker. 
For each reproduced report, \data includes the original bug report, an \texttt{\small apk} file for the buggy version of the app, an executable reproduction script, and metadata regarding the quality of the reproduction steps associated with the original report.
We believe that the \data dataset can be used to facilitate research in automatically analyzing, understanding, reproducing, localizing, and fixing bugs for mobile applications as well as other software maintenance activities more broadly.
 Links to Dataset: \texttt{\small\url{https://doi.org/10.5281/zenodo.4646313}} \\
\texttt{\small\url{https://github.com/SageSELab/AndroR2}}
\end{abstract}

%\begin{IEEEkeywords}
%\end{IEEEkeywords}

\section{Introduction}
\label{sec:intro}

Software maintenance activities are known to be generally time consuming, so much so that prior studies have illustrated they can often comprise more than half of the development effort for a given software project~\cite{2008_tassey_testing}. While developers  
carry out a vast array of maintenance activities, 
perhaps no artifact is more central to a wider variety of maintenance tasks than the bug reports filed in issue tracking systems. 

A sizable portion of the research around automating activities related to bug report management has focused upon the domain of mobile applications~\cite{2019_fse_chaparro_assessing,2018_issta_fazzini_automatically,2019_icse_yu_recdroid,2015_fse_moran_auto}, as mobile devices and apps continue to grow in their ubiquity and popularity (e.g., 3 million apps on Google Play~\cite{2021_google_play}), and developers require new techniques and tools to cope with challenges related to rapidly evolving and fragmented devices~\cite{2012_han_wcre_fragmentation,android-fragmentation} and a growing user base~\cite{2014_jones_tech_report}. 
However, most existing studies on automating bug management activities for mobile applications are limited by the lack of a reliable and systematically created dataset, and tend to be evaluated against a manually curated sets of bug reports~\cite{2019_fse_chaparro_assessing,2018_issta_fazzini_automatically,2019_icse_yu_recdroid}. 
This complicates measuring the progress of automation in this research area due to the difficulty to compare techniques developed for similar maintenance tasks.

To help address these current limitations of research on automated management of mobile app bugs, this paper introduces the \data dataset~\cite{appendix}. \data consists of 90 manually-reproduced bug reports, collected via an in-depth analysis of 459 issues systematically mined from the bug tracking system of Android apps hosted on GitHub and available on the Google Play Store~\cite{2021_google_play}. \data includes 23 reports describing a failure that manifests as a crash and 67 reports detailing a non-crashing failure. Each bug report in the \data dataset was manually verified to be fully reproducible by at least two authors 
of this paper; the dataset includes both executable \texttt{.apk} files and Android device configurations that allow for the reproduction of each reported bug.  
Furthermore, given that prior work illustrated the criticality of reproduction steps (S2Rs) to the quality of bug reports~\cite{2010_zimmerman_tse_good_reports,2019_fse_chaparro_assessing}, for each included bug, we offer an extensive analysis of the S2Rs  and include structured data representation of information related to the number of steps, issues with the reported steps (e.g., missing information), as well as the setup or environmental constraints required for the bug to manifest. 
Finally, we include automated scripts that fully reproduce the failures described in the reports.

We believe that this dataset will support future work and studies related to automating various bug reporting activities. In particular, by including both the (often) flawed user reported S2Rs, as well as "ground truth" sets of reproduction steps, we believe that the dataset can directly support future work on bug report quality assessment and reproduction. Additionally, we describe how \data and its intermediate artifacts can be used and extended to support research in a broad range of software testing and maintenance activities.

\section{Originality of the Dataset}
\label{sec:related}

There are a number of datasets from prior work on automated bug report management. Most notable are the datasets from the papers introducing {\sc Fusion}~\cite{2015_fse_moran_auto} (15 reports), {\sc Euler}~\cite{2019_fse_chaparro_assessing} (24 reports), {\sc Yakusu}~\cite{2018_issta_fazzini_automatically} (48 reports and 12 trivial reports), and ReCDroid~\cite{2019_icse_yu_recdroid} (51 reports).

The \data dataset is largely \textit{complimentary} to these datasets (it includes only one bug report contained in these datasets) 
and exhibits several key differences that set it apart.  First, the \data dataset includes different types of bugs representing a more diverse population of faults (\ie faults leading to crashing and non-crashing failures), whereas the largest datasets from prior work focused \textit{exclusively on crashes}~\cite{2018_issta_fazzini_automatically,2019_icse_yu_recdroid}. Second, \data was built using a systematic methodology for mining, filtering, and reproducing the reports that focuses on reports written by non-contributors (e.g. likely originating from end-users). This diversity in reporters is important for future work on automatically analyzing bug reports, as prior work has shown that reporters of different levels of expertise construct bug reports differently~\cite{2016_jws_rodeghero_an}. 
Third, \data contains important metadata related to the S2Rs of the collected bugs, which can support new and continuing lines of research as described in Section~\ref{sec:usage}. Finally, our dataset is both the largest to date and contains automated
 scripts for reproducing the reported bugs.
\section{Dataset Creation}
\label{sec:methodology}

\begin{figure*}[t!]
	\centering
	\vspace{-1em}
	\centerline{\includegraphics[width=0.8\textwidth]{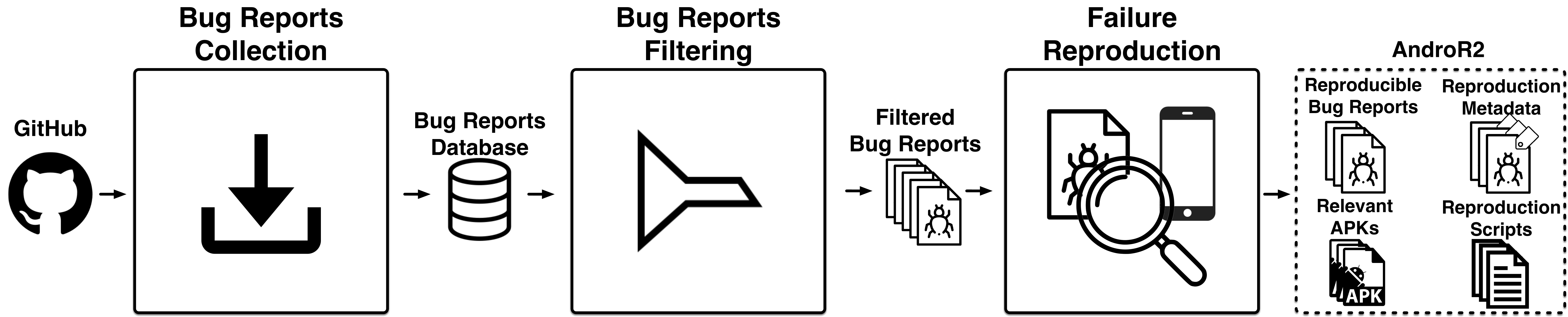}}
	%\vspace{-5pt}
	\caption{High-level overview on the methodology used to build \data.} \label{fig:methodology}
	\vspace{-16pt}
\end{figure*}

In this section, we present the methodology used to build \data and also discuss the key challenges faced during its construction. Figure~\ref{fig:methodology} provides a high-level overview of our methodology workflow, which consisted of three main phases: the \textit{bug reports collection} phase, the \textit{bug reports filtering} phase, and the \textit{failure reproduction} phase. The rest of this section describes the three phases in detail.

\subsection{Bug Reports Collection}
In the \textit{bug reports collection} phase, we built a dataset of bug reports based on GitHub issues~\cite{2021_github_issues}. We selected GitHub as the source for building \data, both due to its popularity and integration of source code hosting and issue tracking, which allowed us to both mine relevant bug reports and build executable \texttt{.apk} files from source code. To identify relevant issues, we built a tool based on the GitHub REST API~\cite{2021_github_rest_api} that mines issues containing reported bugs. 

First, we mined all GitHub issues posted between January 1st, 2016 and November 1st, 2020 that met the following criteria: (i) are part of repositories that use Java and (ii) have the label ``bug''. We performed this task by creating one query for each day during the 5 year date range,  and the query was retrieving the issues created on the day (each individual query never returned more than 1,000 results, which is the 
search limit associated with the GitHub REST API~\cite{2021_github_search}). 
We considered issues created in the last five years to avoid frequently encountered problems in building or finding the executables (\ie \texttt{.apk}s~\cite{2021_google_build}) of app versions corresponding to older reports.  
We selected issues labelled as ``bug'' to effectively identify those likely related to bug reports and ignore issues discussing ideas, enhancements, or tasks.

Second, for each issue identified in the first step, we used our mining tool to determine whether the issue was part of a repository containing an Android app. This was accomplished by cloning the repository associated with the issue and checking  
whether it contained an \texttt{\small AndroidManifest.xml} file, as each Android app requires this file to properly compile~\cite{2021_app_manifest_overview}. Once we confirmed that an issue was part of a repository likely containing an Android app, we included the issue as a document in a MongoDB~\cite{2021_mongodb} database. 
At the end of this first phase, this \emph{bug reports database} contained 82,455 issues.

\subsection{Bug Reports Filtering}
In this phase, we systemically collected a set of issues from the \textit{bug reports database} with the objective of building a representative set of bug reports which could then be further analyzed manually. 
To this end, we first identified and selected issues that belong to repositories whose app is on the Google Play Store to help eliminate issues with trivial apps. This resulted in a set of 28,501 issues. 
Second, because \data aims to foster research on bug reports and their S2Rs, we collected issues that contain the word `steps' in any portion of the report. We used this measure to avoid manually processing a large number of issues without S2Rs during the failure reproduction phase. This step resulted in 6,365 issues. 
Third, we further refined the set of issues to only contain those created by a GitHub user that had \textit{not} contributed to the repository, resulting in 3,842 issues. 
Fourth, we selected issues that were closed at the time the issues were mined (November 2020). 
We focused on closed issues so that we could more easily identify whether the issues were also originally reproduced by the developers and to provide a higher likelihood of manual bug reproduction in the next phase. This filtering resulted in 3,005 reports.
Fifth, after analyzing the set of issues, we found that some repositories had a much larger number of issues compared to others. To avoid overfitting \data to a specific app, we considered at most ten issues per repository. When a repository had more than ten issues, we randomly selected ten from this set (we did this operation for 24 repositories). The resulting set of issues consisted of 459 bug reports for 121 apps.
Finally, to further facilitate the process of reproducing the issues,  two authors of this paper read each of the 459 issues and filtered out those that were either not reproduced by the developers (170) by looking for this information in the discussion associated with the report or were trivial, \ie occur by simply opening the app (15). This resulted in 274 issues for 88 apps, and we call this set the \textit{filtered bug reports}.

\subsection{Failure Reproduction Phase}
\vspace{-0.5em}
In the last phase of our dataset creation process, we manually processed the filtered 274 bug reports to create \data. Specifically, we first analyzed each of the filtered bug reports to manually reproduced the failures. 
This process resulted in 90 successfully \textit{reproducible bug reports}.
Then, for each reproducible report, we derived \textit{reproduction metadata} (detailed in Section~\ref{sec:dataset-c}) on the quality of the S2Rs in bug reports, and created executable \textit{reproduction scripts}. The reproducible bug reports, the reproduction metadata, the reproduction scripts, and the \textit{relevant APKs} (\ie the relevant app executables) make up the content of \data dataset. 

Five of the authors worked to reproduce the failures described in the filtered bug reports. 
When reproducing a bug report, the authors first identified the version of the app associated with that report. 
If the bug report did not provide the version information, the authors used the latest version of the app that was released before the date the report was submitted. Second, the authors checked whether the report identified the version of the Android OS on which the user experienced the failure. If the report contained this information, the authors used that version to reproduce the failure. Otherwise, the authors extracted the \texttt{\small targetSdkVersion} value~\cite{2021_google_targetsdk} from the app and used that value as the version of the Android OS on which to reproduce the failure. Finally, the authors attempted to reproduce the failure by interacting with a Pixel 2 emulator running the app and Android versions they identified.

To reproduce a failure, the authors followed the S2Rs contained in the bug report by mapping the steps to GUI actions in the app. If a report had missing S2Rs, the authors manually explored the functionality of the app to identify the minimal sequence of GUI actions that would account for those missing steps. (The authors used a trial-and-error approach in this situation.) For successfully reproduced failures, 
the authors repeated the reproduction steps at least one additional time as a sanity check and two authors tried to reproduced the same bug report to ensure that reproduced failure was the same as the one described in the report.  The authors then encoded the GUI actions in a reproduction script using the UIAutomator framework. 

To ensure the reliability of the manual reproduction phase, the work was divided into two stages. In the first stage, we divided the filtered bug reports among the five authors, and they attempted to reproduce the bug reports. In the second stage, we selected the bug reports that were not reproduced successfully and reassigned them so that another author made a second reproduction attempt. At the end of the second stage, we had 90 reproducible bug reports. The reasons we could not reproduce certain bug reports were as follows: we could not reproduce the failure even if we followed the S2Rs in the report (76 cases); could not build or find a suitable APK for reproducing the bug (56 cases); the bug report required the use of an additional device (24 cases); a personal account is needed for reproducing the report (16 cases);
the bug report required the use of additional files we could not access (7 cases); the bug report required the use of a real device (5 cases). It is worth noting that the effort required to undertake this process was quite high: around six man-months. This phase was the one that introduced the main challenges in building \data as it required carefully mapping S2Rs to GUI actions while accounting for missing or ambiguous S2Rs.
\section{Dataset Characteristics}
\label{sec:dataset-c}

\begin{figure*}[t!]
\centering
\scriptsize
\vspace{-1.5em}
\begin{minipage}[t]{0.32\textwidth}
%\textbf{Title:} Deposit / Withdrawal change existing entry
\begin{tcolorbox}[colback=white, boxsep=0pt, left=2pt, right=2pt, before skip=2pt,after skip=0pt]

%\begin{tcolorbox}[colback=white, boxsep=0pt, left=4pt, right=4pt, before skip=4pt,after skip=0pt]
\textbf{Title:} Problem with the eraser tab\\
\\
\textbf{Reproduction Steps}\\
1. Click on whiteboard tab while reviewing flashcards and use the feature to write/draw anything\\
2. Click on the erase tab till you have undone the drawing and the erase tab grays out\\
3. Redraw anything by clicking on the whiteboard icon\\
\\
\textbf{Expected Result}\\
Upon redrawing, the eraser tab should revert its color from gray to white\\
\\
\textbf{Actual Result}\\
On redrawing, the eraser tab still remains grayed out.\\
\\
\textbf{Debug info}\\
...AnkiDroid Version = 2.10beta3\\
Android Version = 10...
\vspace{3pt}
\end{tcolorbox}
\caption{Bug report \#160.}
\label{fig:bugreport}
\end{minipage}
\begin{minipage}[t]{0.24\textwidth}
\begin{tcolorbox}[colback=white, boxsep=0pt, left=2pt, right=2pt, before skip=2pt,after skip=0pt]
%\inputminted[
%firstnumber=1,
%stepnumber=1,
%fontsize=\footnotesize,
%numbersep=2pt,
%xleftmargin=5pt
%]
%{json}{code/metadata.json}
%\begin{verbatim}

%\end{verbatim}
\begin{lstlisting}[language=json,basicstyle=\scriptsize\ttfamily,numberstyle=\scriptsize\color{black},aboveskip=0pt,belowskip=0pt]
{"id": 160,
 "github_issue": "https://...",
 "commit_id": "abd1db2...",
 "android_os":"11",
 "failure_type":"gui",
 "s2rs":3,
 "gui_actions":14,
 "setup_gui_actions":8,
 "rs_gui_actions":6,
 "rs_missing_gui_actions":0,
 "multiple_gui_action_s2rs:":1,
 "gui_actions_in_mgas2rs":4,
 "sgas_in_android_os":1,
 "sgas_outside_app":0,
 "rsgas_in_android_os":0,
 "rsgas_outside_app":0,
 "sga_nl_description": "...",
 "rsga_nl_description": "..."}
\end{lstlisting}
\vspace{8pt}
\end{tcolorbox}
\caption{Metadata.}
\label{fig:metadata}
\end{minipage}
\begin{minipage}[t]{0.43\textwidth}
 \begin{tcolorbox}[colback=white, boxsep=0pt, left=2pt, right=2pt, before skip=2pt,after skip=0pt]
\begin{lstlisting}[language=Java,basicstyle=\scriptsize\ttfamily,numberstyle=\scriptsize\color{black},xleftmargin=10pt,aboveskip=0pt,belowskip=0pt]
public class Script160 {
 ...
 @Test
 public void reproduce() {
  ...
  UiObject2 Default = mDevice.wait(
   Until.findObject(By.text( "Default")),2000);
  Default.click();

  UiObject2 More = mDevice.wait(
   Until.findObject(By.desc( "More options")),2000);
  More.click();

  UiObject2 Whiteboard = mDevice.wait(
   Until.findObject(By.text( "Enable whiteboard")),2000);
  Whiteboard.click();

  mDevice.drag(300,300, 600, 600, 1);
  ... }}
\end{lstlisting}
%\vspace{6pt}
\end{tcolorbox}
\caption{Reproduction script.}
\label{fig:script}
\end{minipage}
\vspace{-18pt}
\end{figure*}

\data is available as on Zenodo~\cite{appendix} \& GitHub~\cite{github}. The dataset contains the reproducible bug reports, the reproduction metadata, the relevant \texttt{.apk}s, and the reproduction scripts. The reproducible bug reports are located in the \texttt{reports} folder and are in their original HTML format. The reproduction metadata is in the \texttt{metadata} folder and encoded in JSON files. The relevant APKs are in the \texttt{apks} folder. The reproduction scripts are in the \texttt{scripts} folder and are encoded as UIAutomator tests. We used UIAutomator as it allows for interacting with both the Android OS and other apps, which is necessary for some reports. The dataset also contains a \texttt{README.md} that describes its content. 
Finally, the scripts and tools we developed to build \data are in the \texttt{code} folder.

\vspace{0.05in}
\noindent
{\bf Dataset Details.}
To detail the data stored in \data, we use a bug report ($\mathit{BR}$\textit{\#}$160$) contained in the dataset as a running example. Figure~\ref{fig:bugreport} shows the relevant portion of the bug report, Figure~\ref{fig:metadata} presents the metadata associated with the bug report, and Figure~\ref{fig:script} illustrates part of the reports' reproduction script. 

The bug report contains three S2Rs (under the header ``\textit{Reproduction Steps}'').
The reproduction metadata provides information on the failure reproduction task and how it connects to the bug report's information. The metadata also contains a detail natural language description of all the S2Rs necessary to reproduce the failure of the report. (We provide a full description of the metadata in~\cite{appendix}.)
In short, 
\texttt{id} is the identifier for the bug report in \data,
\texttt{github\_issue} provides the link to the GitHub issue,
and \texttt{failure\_type} describes the type of failure associated with the report. We classified failures under three categories: \textit{crash} identifies a crash in the app, \texttt{output} represents an error in the output of the app, and \texttt{gui} identifies an error in the properties of the app's GUI. 
For $\mathit{BR}$\textit{\#}$160$, the failure is of type \texttt{gui} because an app icon is not displayed correctly. 

Further, \texttt{s2rs} provides the number of S2Rs in the report. We counted S2Rs as follows. If the report provided a bulleted (numbered) list for the S2Rs, we counted how many bulleted (numbered) items are listed. If the S2Rs are described through text paragraphs, we counted the number of sentences in the paragraphs. 
\texttt{gui\_actions} is the number of GUI actions exercised by authors to reproduce the failure. \texttt{setup\_gui\_actions} is the number of GUI actions that were necessary before the author could perform the GUI action(s) associated with the first S2R. 
For $\mathit{BR}$\textit{\#}$160$, there are eight such actions. Two actions are necessary after a fresh install of the app, and six actions are required to create a flashcard required by the S2Rs of the report. \texttt{rs\_gui\_actions} provides the number of report-specific GUI actions, that is, the number of actions associated with the sequence of S2Rs in the report.  
\data also includes scripts for reproducing the failures of the reports. Figure~\ref{fig:script} shows part of the script for reproducing $\mathit{BR}$\textit{\#}$160$ and includes four GUI actions (lines 6, 10, 14, and 18).

\vspace{0.05in}
\noindent
{\bf Dataset Summary Statistics.}
\data contains 90 reproducible bug reports. Among the bug reports, 23 describe crashes, 34 detail output errors, 33 report GUI errors, 77 require setup GUI actions, 33 have report-specific GUI actions not mentioned by the S2Rs, and 57 contain at least one S2R that leads to multiple GUI actions. \data also contains 90 executable scripts reproducing the failure of the bug reports.

\section{Applications, Limitations, and Extensions}
\label{sec:usage}

The primary use of our dataset is to facilitate research related to automated analysis, understanding, and reproduction of bug reports.
For example, recent approaches that utilize program analysis and natural language processing to
automatically reproduce crashes in Android apps~\cite{2016_icst_moran, 2018_issta_fazzini_automatically, zhao:recdroid:ICSE:2019}
could use our dataset to measure the fraction of successfully reproduced reports.
Such approaches can also utilize our manually-produced ``ground truth'' execution scripts, comparing the number of steps in the automated and manually produced versions to ensure that
an automated reproduction does not produce excessively long and cumbersome scripts. \data could be also used be used to perform bug report prioritization based on the quality of S2Rs.  Finally, the dataset could be used by techniques that aim to map S2Rs into GUI actions~\cite{2020_issta_liu_automated}.

\vspace{0.02in}
\noindent
{\bf Additional Usages.}
Beyond the bug report management scenarios, \data could be used to benefit research in:

\vspace{0.02in}
\noindent
1) Targeted app exploration approaches, e.g.,~\cite{2012_zheng_smartdroid,2013_issta_collider,2015_sschutte_condroid,2019_ase_lai_goalexplorer},
rely on static and dynamic program analysis to force execution towards a particular line of code or app screen.
Such approaches could set faults in our dataset as exploration targets,
verifying the ability of an approach to successfully reach the target (without relying on the bug report for that purpose) and,
when successful, comparing the number of steps in the produced execution to those in our execution scripts.

\vspace{0.02in}
\noindent
2) \data can be used as a benchmark to assess the efficiency and scalability of dynamic application slicing techniques~\cite{2019_icse_azim_androidslicer, 2021_icst_ahmed_mandoline}. Such techniques are
typically used for debugging purposes, with a fault set as the slicing criteria.

\vspace{0.02in}
\noindent
3) Similarly, fault localization techniques~\cite{2016_tse_wong_localization}, especially for Android applications~\cite{2015_mobilesoft_mirzaei_localization}, could benefit from
our collection of apps with ``verified'' faulty behaviors.

\vspace{0.02in}
\noindent
4) Finally, automated approaches for analyzing video recordings of Android app usages into replayable scenarios, e.g.,~\cite{2020_icse_bernal}, could use execution scripts in our dataset as a testbed for evaluating whether the tools can accurately analyze and replay scenarios from screen recordings.

\vspace{0.05in}
\noindent
{\bf Dataset Extensions.}
The most obvious way of extending our dataset is to analyze and reproduce more bug reports. 
To facilitate such extension,
we include with \data a tool for mining, extracting, and storing issues from GitHub. 
In addition, we provide a database of current 82,455 collected issues. 

\data can be further extended through the addition of assertion statements to the reproduction scripts of bugs without an explicit oracle (i.e., crashing bugs) in order to create a set of functional tests.
With such assertions added, the dataset could help to foster research related to 
automated bug repair techniques, to validate that the proposed repairs are successful and lead to a passing test.
Furthermore, augmenting the execution scripts with assertions can provide the needed reference model to evaluate
automated assertion generation techniques, e.g.,~\cite{2020_watson_icse_assertions}.

Another way to extend \data is by including multiple different variants of bug reproduction scripts,
i.e., those that perform different sets of actions that lead to the same bug, as opposed to the minimal action sequences currently provided.
Such an extension could facilitate research which examines test case selection and prioritization techniques~\cite{2017_compsurv_kazmi_testselection}
as well as research on detecting duplicate video-based bug reports~\cite{2021_icse_cooper}.

\section{Conclusions}
\label{sec:conclusions}

This paper presented the \data dataset. We believe that this dataset fosters open, reproducible future research on automated bug report management, software testing, and software maintenance more broadly. 

\section*{Acknowledgment}

This work was partially supported by a gift from Facebook  and the NSF CCF-1955853 grant. 
%Any opinions, findings, and conclusions expressed herein are the authors' and do not necessarily reflect those of the sponsors.

\balance
\bibliographystyle{IEEEtran}
\bibliography{00}

% Generated by IEEEtran.bst, version: 1.14 (2015/08/26)
\begin{thebibliography}{10}
\providecommand{\url}[1]{#1}
\csname url@samestyle\endcsname
\providecommand{\newblock}{\relax}
\providecommand{\bibinfo}[2]{#2}
\providecommand{\BIBentrySTDinterwordspacing}{\spaceskip=0pt\relax}
\providecommand{\BIBentryALTinterwordstretchfactor}{4}
\providecommand{\BIBentryALTinterwordspacing}{\spaceskip=\fontdimen2\font plus
\BIBentryALTinterwordstretchfactor\fontdimen3\font minus
  \fontdimen4\font\relax}
\providecommand{\BIBforeignlanguage}[2]{{%
\expandafter\ifx\csname l@#1\endcsname\relax
\typeout{** WARNING: IEEEtran.bst: No hyphenation pattern has been}%
\typeout{** loaded for the language `#1'. Using the pattern for}%
\typeout{** the default language instead.}%
\else
\language=\csname l@#1\endcsname
\fi
#2}}
\providecommand{\BIBdecl}{\relax}
\BIBdecl

\bibitem{2008_tassey_testing}
G.~Tassey, ``The economic impacts of inadequate infrastructure for software
  testing,'' National Institute of Standards and Technology, Tech. Rep., 2002.

\bibitem{2019_fse_chaparro_assessing}
O.~Chaparro, C.~Bernal-C\'{a}rdenas, J.~Lu, K.~Moran, A.~Marcus, M.~Di~Penta,
  D.~Poshyvanyk, and V.~Ng, ``{Assessing the Quality of the Steps to Reproduce
  in Bug Reports},'' in \emph{Proceedings of the 2019 27th ACM Joint Meeting on
  European Software Engineering Conference and Symposium on the Foundations of
  Software Engineering}.\hskip 1em plus 0.5em minus 0.4em\relax Association for
  Computing Machinery, 2019, p. 86–96.

\bibitem{2018_issta_fazzini_automatically}
M.~Fazzini, M.~Prammer, M.~d'Amorim, and A.~Orso, ``{Automatically Translating
  Bug Reports into Test Cases for Mobile Apps},'' in \emph{Proceedings of the
  ACM SIGSOFT International Symposium on Software Testing and Analysis}.\hskip
  1em plus 0.5em minus 0.4em\relax Association for Computing Machinery, 2018,
  pp. 141--152.

\bibitem{2019_icse_yu_recdroid}
Y.~Zhao, T.~Yu, T.~Su, Y.~Liu, W.~Zheng, J.~Zhang, and W.~G.~J. Halfond,
  ``{ReCDroid: Automatically Reproducing Android Application Crashes from Bug
  Reports},'' in \emph{Proceedings of the 41st International Conference on
  Software Engineering}.\hskip 1em plus 0.5em minus 0.4em\relax IEEE Press,
  2019, pp. 128–--139.

\bibitem{2015_fse_moran_auto}
K.~Moran, M.~Linares-V\'{a}squez, C.~Bernal-C\'{a}rdenas, and D.~Poshyvanyk,
  ``{Auto-Completing Bug Reports for Android Applications},'' in
  \emph{Proceedings of the 2015 10th Joint Meeting on Foundations of Software
  Engineering}.\hskip 1em plus 0.5em minus 0.4em\relax Association for
  Computing Machinery, 2015, pp. 673--–686.

\bibitem{2021_google_play}
\BIBentryALTinterwordspacing
(2021, Jan.) Google play. [Online]. Available:
  \url{https://play.google.com/store}
\BIBentrySTDinterwordspacing

\bibitem{2012_han_wcre_fragmentation}
\BIBentryALTinterwordspacing
D.~Han, C.~Zhang, X.~Fan, A.~Hindle, K.~Wong, and E.~Stroulia, ``Understanding
  android fragmentation with topic analysis of vendor-specific bugs,'' in
  \emph{Proceedings of the 2012 19th Working Conference on Reverse
  Engineering}, ser. WCRE '12.\hskip 1em plus 0.5em minus 0.4em\relax
  Washington, DC, USA: IEEE Computer Society, 2012, pp. 83--92. [Online].
  Available: \url{http://dx.doi.org/10.1109/WCRE.2012.18}
\BIBentrySTDinterwordspacing

\bibitem{android-fragmentation}
``Android fragmentation statistics
  \url{http://opensignal.com/reports/2014/android-fragmentation/},'' 2014.

\bibitem{2014_jones_tech_report}
N.~Jones, ``Seven best practices for optimizing mobile testing efforts,''
  Gartner, Technical Report G00248240.

\bibitem{appendix}
\BIBentryALTinterwordspacing
(2021, Mar.) {{AndroR2 Dataset}}. [Online]. Available:
  \url{https://doi.org/10.5281/zenodo.4646313}
\BIBentrySTDinterwordspacing

\bibitem{2010_zimmerman_tse_good_reports}
T.~{Zimmermann}, R.~{Premraj}, N.~{Bettenburg}, S.~{Just}, A.~{Schroter}, and
  C.~{Weiss}, ``What makes a good bug report?'' \emph{IEEE Transactions on
  Software Engineering}, vol.~36, no.~5, pp. 618--643, 2010.

\bibitem{2016_jws_rodeghero_an}
P.~Rodeghero, D.~Huo, T.~Ding, C.~McMillan, and M.~Gethers, ``An empirical
  study on how expert knowledge affects bug reports,'' p. 542–564, 2016.

\bibitem{2021_github_issues}
\BIBentryALTinterwordspacing
(2021, Jan.) About issues. [Online]. Available:
  \url{https://docs.github.com/en/github/managing-your-work-on-github/about-issues}
\BIBentrySTDinterwordspacing

\bibitem{2021_github_rest_api}
\BIBentryALTinterwordspacing
(2021, Jan.) Github rest api. [Online]. Available:
  \url{https://docs.github.com/en/rest}
\BIBentrySTDinterwordspacing

\bibitem{2021_github_search}
\BIBentryALTinterwordspacing
(2021, Jan.) Search. [Online]. Available:
  \url{https://docs.github.com/en/rest/reference/search}
\BIBentrySTDinterwordspacing

\bibitem{2021_google_build}
\BIBentryALTinterwordspacing
(2021, Jan.) Build and run your app. [Online]. Available:
  \url{https://developer.android.com/studio/run}
\BIBentrySTDinterwordspacing

\bibitem{2021_app_manifest_overview}
\BIBentryALTinterwordspacing
(2021, Jan.) App manifest overview. [Online]. Available:
  \url{https://developer.android.com/guide/topics/manifest/manifest-intro}
\BIBentrySTDinterwordspacing

\bibitem{2021_mongodb}
\BIBentryALTinterwordspacing
(2021, Jan.) The database for modern applications. [Online]. Available:
  \url{https://www.mongodb.com}
\BIBentrySTDinterwordspacing

\bibitem{2021_google_targetsdk}
\BIBentryALTinterwordspacing
(2021, Jan.) <uses-sdk>. [Online]. Available:
  \url{https://developer.android.com/guide/topics/manifest/uses-sdk-element}
\BIBentrySTDinterwordspacing

\bibitem{github}
\BIBentryALTinterwordspacing
(2021, Mar.) {{AndroR2 Dataset}}. [Online]. Available:
  \url{https://github.com/SageSELab/AndroR2}
\BIBentrySTDinterwordspacing

\bibitem{2016_icst_moran}
K.~Moran, L.-V. Mario, C.~Bernal-C\'{a}rdenas, C.~Vendome, and D.~Poshyvanyk,
  ``{Automatically Discovering, Reporting and Reproducing Android Application
  Crashes},'' in \emph{Proceedings of the IEEE International Conference on
  Software Testing, Verification and Validation}, 2016, pp. 33--44.

\bibitem{zhao:recdroid:ICSE:2019}
Y.~Zhao, T.~Yu, T.~Su, Y.~Liu, W.~Zheng, J.~Zhang, and W.~G.~J. Halfond,
  ``{ReCDroid: Automatically Reproducing Android Application Crashes from Bug
  Reports},'' in \emph{Proceedings of the ACM/IEEE International Conference on
  Software Engineering}, 2019, pp. 128--139.

\bibitem{2020_issta_liu_automated}
H.~Liu, M.~Shen, J.~Jin, and Y.~Jiang, ``Automated classification of actions in
  bug reports of mobile apps,'' in \emph{Proceedings of the 29th ACM SIGSOFT
  International Symposium on Software Testing and Analysis}.\hskip 1em plus
  0.5em minus 0.4em\relax Association for Computing Machinery, 2020, p.
  128–140.

\bibitem{2012_zheng_smartdroid}
C.~Zheng, S.~Zhu, S.~Dai, G.~Gu, X.~Gong, X.~Han, and W.~Zou, ``{SmartDroid: An
  Automatic System for Revealing UI-based Trigger Conditions in Android
  Applications},'' in \emph{Proceedings of the Second ACM Workshop on Security
  and Privacy in Smartphones and Mobile Devices}, 2012, pp. 93--104.

\bibitem{2013_issta_collider}
C.~S. Jensen, M.~R. Prasad, and A.~M{\o}ller, ``{Automated Testing with
  Targeted Event Sequence Generation},'' in \emph{Proceedings of the ACM
  SIGSOFT International Symposium on Software Testing and Analysis}, 2013, pp.
  67--77.

\bibitem{2015_sschutte_condroid}
J.~Sch\"{u}tte, R.~Fedler, and D.~Titze, ``{ConDroid: Targeted Dynamic Analysis
  of Android Applications},'' in \emph{Proceedings of IEEE International
  Conference on Advanced Information Networking and Applications}, 2015, pp.
  571--578.

\bibitem{2019_ase_lai_goalexplorer}
D.~Lai and J.~Rubin, ``{Goal-Driven Exploration for Android Applications},'' in
  \emph{Proceedings of the IEEE/ACM International Conference on Automated
  Software Engineering}, 2019, pp. 115--127.

\bibitem{2019_icse_azim_androidslicer}
T.~Azim, A.~Alavi, I.~Neamtiu, and R.~Gupta, ``{Dynamic Slicing for Android},''
  in \emph{Proceedings of the ACM/IEEE International Conference on Software
  Engineering}, 2019, pp. 1154--1164.

\bibitem{2021_icst_ahmed_mandoline}
K.~Ahmed, M.~Lis, and J.~Rubin, ``{Mandoline: Dynamic Slicing of Android
  Applications with Trace-Based Alias Analysis},'' in \emph{Proceedings of the
  IEEE International Conference on Software Testing, Verification and
  Validation}, 2021.

\bibitem{2016_tse_wong_localization}
W.~E. Wong, R.~Gao, Y.~Li, R.~Abreu, and F.~Wotawa, ``{A Survey on Software
  Fault Localization},'' \emph{{IEEE} Trans. Software Eng.}, vol.~42, no.~8,
  pp. 707--740, 2016.

\bibitem{2015_mobilesoft_mirzaei_localization}
H.~Mirzaei and A.~Heydarnoori, ``{Exception Fault Localization in Android
  Applications},'' in \emph{Proceedings of the {ACM} International Conference
  on Mobile Software Engineering and Systems}, 2015, pp. 156--157.

\bibitem{2020_icse_bernal}
C.~Bernal-C\'{a}rdenas, N.~Cooper, K.~Moran, O.~Chaparro, A.~Marcus, and
  D.~Poshyvanyk, ``{Translating Video Recordings of Mobile App Usages into
  Replayable Scenarios},'' in \emph{Proceedings of the ACM/IEEE International
  Conference on Software Engineering}, 2020, pp. 309–--321.

\bibitem{2020_watson_icse_assertions}
C.~Watson, M.~Tufano, K.~Moran, G.~Bavota, and D.~Poshyvanyk, ``{On Learning
  Meaningful Assert Statements for Unit Test Cases},'' in \emph{Proceedings of
  International Conference on Software Engineering}, 2020, pp. 1398--1409.

\bibitem{2017_compsurv_kazmi_testselection}
R.~Kazmi, D.~N.~A. Jawawi, R.~Mohamad, and I.~Ghani, ``{Effective Regression
  Test Case Selection: A Systematic Literature Review},'' \emph{ACM Comput.
  Surv.}, vol.~50, no.~2, 2017.

\bibitem{2021_icse_cooper}
N.~Cooper, C.~Bernal-C\'{a}rdenas, O.~Chaparro, K.~Moran, and D.~Poshyvanyk,
  ``{It Takes Two to Tango: Combining Visual and Textual Information for
  Detecting Duplicate Video-Based Bug Reports},'' in \emph{Proceedings of the
  ACM/IEEE International Conference on Software Engineering}, 2021.

\end{thebibliography}

\end{document}